# Investigation of Noise Limitation Standardization and Evaluating the Low Frequency Noise's Influence on Human Performance using Online Psychoacoustic Test

Xuhao Du, Jia Ma and Zhibin Lin

Key Laboratory of Modern Acoustics and Institute of Acoustics, Nanjing University, Nanjing, 210093, China, duxuhao88@gmail.com, jiama1991@gmail.com, zblin@nju.edu.cn

**ABSTRACT**

A lot of research has been carried out to find out the influence of low frequency noise on human behavior. Based on currently available results, some countries have been developing National Standard to limit the low frequency noise to protect people's health and working performance. However, after investigating 28 National Standards about noise limitation of working machine and other noise emission standards of China, it is found that the stipulation of low frequency noise covers limited aspects. Most of the limitation used A-weight sound level as the evaluation index, which only presents the loudness perception of sound, not its influence. In this paper, two connected topics are presented. One is to investigate the Chinese National Standard and International Standard, in order to summarize all the noise limitation on low frequency noise. The second topic is the application of online psychoacoustic test to evaluate the influence of low frequency noise on human performance, which can enlarge the subjective testing sample size in order to facilitate the establishment of related National Standard. The online Hass effect experiment is repeated to verify the accuracy of online psychoacoustic test. The volunteers are asked to conduct several sessions of thecolour identification problems under different kind of controlled noise. The accuracy and finishing time are taken into consideration to evaluate their performance under different kinds of noise. The results show that low frequency noise does affect humans' working performance and the existing Chinese Noise limitation Standards are inadequate in low frequency noise limitation.

**INTRODUCTION**

Low frequency noise ($N_{LF}$), defined as the noise between 20 Hz to 200 Hz or 250 Hz, has been recognized as a difficult task for engineers to deal with (K. Persson, 1997). as widespread existing in urban environment, it is hard to be controlled. Moreover, it also does special harm to human health and working performance. Numbers of studies show that overexposure in $N_{LF}$ environments may result in temporary or even permanent threshold shift of humans' hearing ability and it takes longer time to recover compared with other noise-caused hearing problem (Mills, J. H., Osguthorpe, J. D., Burdick, C. K., Pattersonet al, 1983). Furthermore, exposure in $N_{LF}$ may also affect the balance, vestibular system and respiratory, which has been documented in laboratory animals and human beings (Birgitta, Peter & R. F. Soames 1996).

Some specific cases have been studied to find out more evidence of $N_{LF}$'s harm. In 1997, by using the building's ventilation noise attached with $N_{LF}$, Persson Waye found that the exposure to lower frequency noise (31.5 Hz to 125 Hz) resulted in lower social orientation and a tendency to lower pleasure to the participants, compared to the ones who were exposed to the mid frequency noise (K. Persson, R.Rylander, S. Benton, et al, 1997). In 2001, he used questionnaires to study the volunteers' sensitivity to $N_{LF}$ (K. Persson, Johanna, Anders, et al, 2001). They also studied subjects' performance, by observing the volunteers' behavior in verbal grammatical reasoning task under different background noise. In this research, the subjects reflected a higher degree of annoyance and their working capacity was impaired when they worked under conditions of $N_{LF}$.

In 2004, a study of a German man's long-time exposure in $N_{LF}$ was conducted by J Feldman (J Feldmann, FA Pitten, 2004). In this case, the $N_{LF}$ was generated by the heating plant. Although the noise measurement report told that the noise in his house meets their national standard, DIN 45680 and DIN 4150, both physiological and psychological effects occurred on the man. Increasing intensity like indisposition, decreasing in performance, sleeping disturbance, suffering headache, ear pressure, crawl paresthesia or shortness of breath are all his symptom. It indicates that the noise emission standard with limiting values of loudness are unable to protect human beings from the harm of $N_{LF}$.

Actually, this is not the only case that the noise meets the standard while people still feel annoyed. $N_{LF}$ may partly contributes to this phenomenon. People noticed that $N_{LF}$ will do particular harm to human beings 70 years ago. Effort was taken far more on linear sound pressure level ($L_{s,L}$). In 1949, Beranek proposed the A-weight curve and A-weight sound pressure level ($L_{s,A}$), which has been made into standard as the noise emmission limitation. In 1981, Blazier studied more than 200 rooms on its noise.The data was used to make the room cretiria (RC) curve, which has been adapted as the design standard (Blazier, 1981). The product of its relative research, like noise criteria curve (NC) and noise rating curve (NR), has also become the national and international standard. According to the investigation of Chinese National Standard (GB) and some other national standards, $N_{LF}$ standard is limited indeed. $L_{s,A}$ has been pointed out in many studies that it is inefficient to predict the annoyance and harm to human beings (Leventhall, G. 2004). Moreover NR and NC criteria also underestimate the influence of $N_{LF}$ (Leventhall, G., Pelmear, P., & Benton, S, 2003). In other words, the attention to influence of $N_{LF}$ is being underestimated and the existing standard is not comprehensive enough.

In this paper, some national and international standards for noise emission will be investigated to show the limitation of application of $N_{LF}$ around the world. The online psychoacoustic test was conducted to investigate the influence of $N_{LF}$ to people's performance by computational methods, in order to provide reference to the revision of related standards.

**THE INVESTIGATION ABOUT THE LOW FREQUENCY NOISE LIMITS STANDARD AROUND THE WORLD**

As mentioned above, many standards adopt $L_{s,A}$, NC curves or their transformations to limit noise emission. Some of them have the limitation of $N_{LF}$. For example, the Polish $N_{LF}$ criteria proposed in 2001 covers 10 Hz to 250 Hz (Mirowska, M. 2001). In 1997, Germany published the German National Standard (DIN:45680) and gave the limitation from 8 Hz to 100 Hz. In Netherland, the $N_{LF}$ limitation for outdoor environment was presented based on the average low frequency hearing thresholds for an otologically unselected population aged 50 – 60, its reference levels is the binaural hearing threshold for 10% of the population (Sloven, 2001). Moreover, Jakobsen proposed the Danish $N_{LF}$ recommadation in 2001 (Jakobsen, 2001). For assessing the indoor $N_{LF}$, SOSFS 1997:7/E was presented as well (Socialstyrelsen-Sweden, 1996). In 2004, the Japanese Ministry of the Environment issued the *Guide to low frequency sound problem solution*. $N_{LF}$ recommadation from 10 Hz to 80 Hz was given (The Japanese Ministry of the environment, 2004). The University of Salford proposed a reference $N_{LF}$ limitation from 10 Hz to 160 Hz in 2005 (A. Moorhouse, D. Waddington, and M. Adams. 2005). Table 1 gathers these limitations with the ISO hearing threshold. It also shows the curves of each country's national standards or recommendations.

**Table** 1**:** The low frequency noise criteria or recommendation in some countries.

| $f$ (Hz) | P (01) (dB) | G (97) (dB) | N (01) (dB) | D (01) (dB) | S (96) (dB) | C (08) (dB) | A (dB) | J (04) (dB) | B (05) (dB) | ISO (dB) |
|---|---|---|---|---|---|---|---|---|---|---|
| 8 | | 103 | | | | | | | | |
| 10 | 80.4 | 95 | | 90.4 | | | | 92 | 92 | |
| 12.5 | 83.4 | 87 | | 93.4 | | | | 88 | 87 | |
| 16 | 66.7 | 79 | | 76.7 | | | | 83 | 83 | |
| 20 | 60.5 | 71 | 74 | 70.5 | | | | 76 | 74 | 74.3 |
| 25 | 54.7 | 63 | 64 | 64.7 | | | | 70 | 64 | 65 |
| 31.5 | 49.3 | 55.5 | 55 | 59.4 | 56 | 69 | | 64 | 56 | 56.3 |
| 40 | 44.6 | 48 | 46 | 54.6 | 49 | | | 57 | 49 | 48.4 |
| 50 | 40.2 | 40.5 | 39 | 50.2 | 43 | | | 52 | 43 | 41.7 |
| 63 | 36.2 | 33.5 | 33 | 46.2 | 41.5 | 51 | | 47 | 42 | 35.5 |
| 80 | 32.5 | 28 | 27 | 42.5 | 40 | | | 41 | 40 | 29.8 |
| 100 | 29.1 | 23.5 | 22 | 39.1 | 38 | | | | 38 | 25.1 |
| 125 | 26.1 | | | 36.1 | 36 | 39 | | | 36 | 20.7 |
| 160 | 23.4 | | | 33.4 | 34 | | | | 34 | 16.8 |
| 200 | 20.9 | | | | 32 | | | | | 13.8 |
| 250 | 18.6 | | | | | 30 | | | | 11.2 |

P: Poland     G: Germany     N: Netherland     D: Denmark     S: Sweden
C: China     A: America     J: Japan     B: British

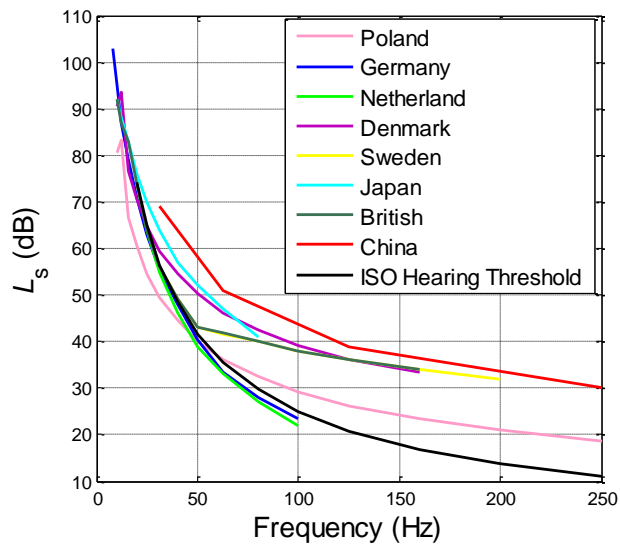

**Figuer 1:** Low Frequency Noise Criteria in some countries

As shown in table 1 and figure 1, it is obvious that the GB have the following flaws: insufficient frequency range and higher $L_{s,L}$ limitation. All other countries mentioned above use 1/3 octave band while in GB, the criteria used octave band. Moverover, the red line shown in figure 1 indicates that the Chinese $N_{LF}$ limitation is higher than any other countries in all bands. After investigating into GB related to noise limitation, only the "GB12348-2008 Emission standard for industrial enterprises noise at boundary" and "GB22337-2008 Emission standard for community noise", have the detailed limitation of $L_{s,A}$ in octave bands from 22~707 Hz, which are issued both in 2008. It is a great improvement based on "GB12348-90 Standard of noise at boundary of industrial enterprises". However, in later issued noise criteria (from 2009 to 2012), the $N_{LF}$ limitations are still lacking (GB16710-2010, GB 24389-2009, GB 24929-2010, GB 26483-2011, GB 26484-2011, GB 28245-2012). Instead, all of them used $L_{s,A}$ as their standards. However, facilities like airplane, motor vehicles and air conditioners can become sources of $N_{LF}$. As a result, no restriction can be applied to manufactories to limit the noise on particular frequency for this kind of products, thus it is difficult to guarantee that sound of noise is small enough on low frequency. While one of the advantages in GB is that it covers a larger range of machine like automatic metalforming machine, hydralic press, mechanical press, accelerating all-terrain vehicles and so on (referred to figure 2).They are also specific to their particles.

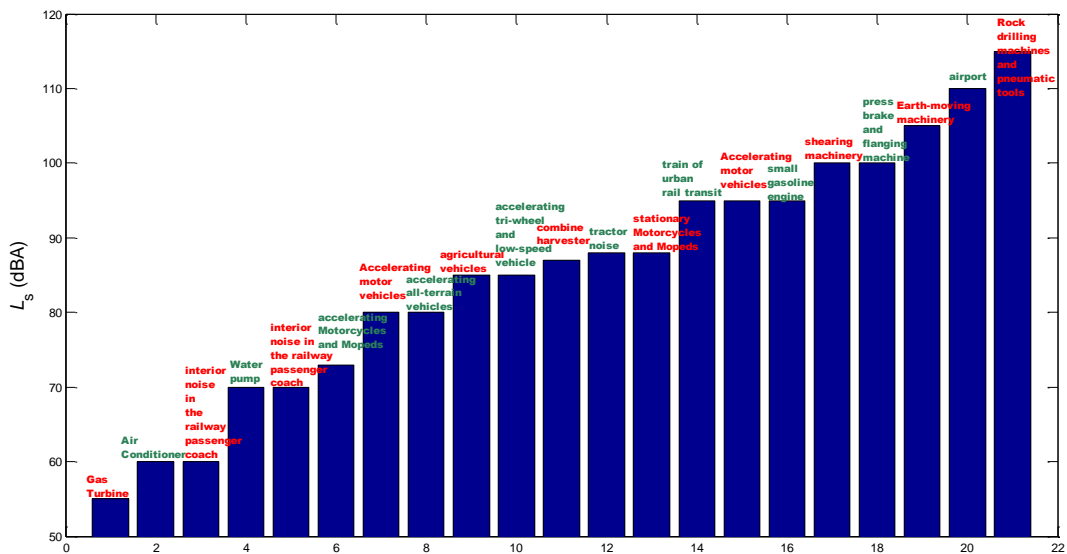

**Figure 2:** Standards for some sources of noise in China. Some of the limitations are average values

## METHOLOGY ABOUT THE ONLINE PSYCHOACOUSTIC TEST

The online psychoacoustic tests are used to investigate the influence of $N_{LF}$ on human working performance with 40 volunteers. Firstly, a special tape (ST) was played back with earphone, according to which the volunteers adjust the volume. Then they cannot change the volume any more. The Hass effect test is following to verify the online psychoacoustic test. Then, 9 different tests with 20 colour recognition problems were given to the volunteers. The first test had not any noise in order to help them get familiar with the operations. Then they had 8 similar tests under different background noise, whose orders were different (random?) among different volunteers. The introduction of the noises was followed in the coming subsection. The accuracies and times were recorded and taken into account as their performance. A Visual Basic program was made to assemble all the procedures and was posted online in order to guide the volunteers to complete the test on internet.

### Hass Effect test

Two identical dial tones were used as the sound tape sources to test the Hass effect. The tape consists of 25 groups and two tones make up one. The internal time between the two tones increases group by group from 10 ms to 70 ms, with the step of 2.5 ms and 1 s interval between the adjoining groups. The volunteers were asked to click a button once they could identify two tones. The duration till they click the button was recorded and the average time would be calculated after the results were gathered.

**Noise**

8 different kinds of noises are used separately in 8 different tests. The first 3 noises are the standard noise, including the pink noise (PN), white noise with the same $L_{s,A}$ with PN (WNA) and another white noise with the same $L_{s,L}$ with PN (WNL). The last 5 noises are based on the recording in a construction site by Sony PCM - D50 with 48000 Hz sample frequency and 24 bits precision, near Nanjing University, Gulou Campus. Each of the 5 noises has at least one different frequency component or $L_s$. The first one is the original constructive noise (CN), the second and the third ones are the original CN with 200 Hz and 500 Hz high pass filters, respectively, of which $L_{s,A}$ are the same with CN (CNA200, CNA500). The fourth and the fifth ones are similar to CNA200 and CNL200. The difference is that they have the same $L_{s,L}$ with CN (CNL200, CNL500). The 1/3 octave spectrum from 25 Hz to 20000 Hz (30 bands) is shown in figure 3, where the '-L' and '-A' represent the $L_{s,L}$ and $L_{s,A}$, respectively. Figure 3(b) shows that the $N_{LF}$ plays an important role in the construction noise. All the 8 noises last for 90 seconds to ensure that the volunteers have enough time to finish their 20 colour recognition tasks.

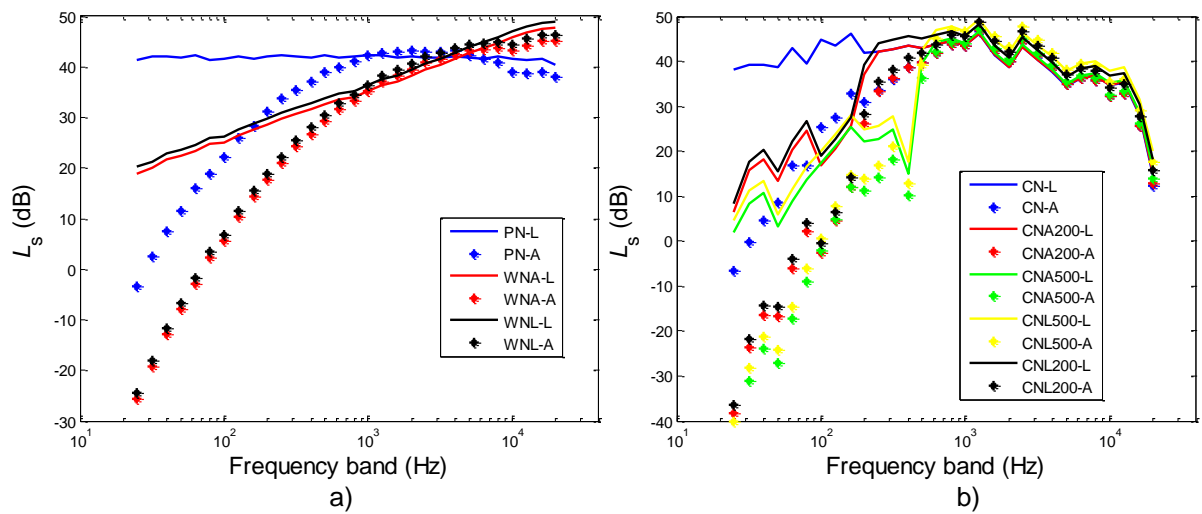

**Figure 3:** The 1/3 octave spectrum of 8 kinds of noise. a) the frequency component of pink noise (PN-L, PN-A) and white noise with the same $L_{s,A}$ (WNA-L, WNA-A) and $L_{s,L}$ (WNL-L, WNL-A) with the pink noise, respectively. b) The frequency component of construction site noise (CN-L, CN-A) and CN under 200 Hz and 500 Hz high pass filter, with the same $L_{s,A}$ (CNA200-L, CNA200-A, CNA500-L, CNA500-A) and $L_{s,L}$ (CNL200-L, CNL200-A, CNL500-L, CNL500-A) with the pink noise, respectively

**The sound pressure level**

A speech tape (ST) was played back by the earphone for the volunteers to fix their volume into a comfortable one. Taking the volume of the tape as the reference $L_s$, the relative $L_s$ of other noises is shown in table 2.

**Colour identification problems**

All the colour identification questions were made up by 3 Chinese characters, red, blue and green, which were in 3 primary colours, red, blue and green randomly. Volunteers were required to choose the characters' colour. Every 20 colour identification questions in 9 lists had the same order of colour and different order of character in order to control the difficulty of each list. The orders of noise were arranged randomly to avoid the influence from it.

**Table 2:** The relative volume of other 8 different noises with the ST.

| No. | Noise | $L_{s,L}$ (dB) | $L_{s,A}$ (dBA) | No. | Noise | $L_{s,L}$ (dB) | $L_{s,A}$ (dBA) |
|---|---|---|---|---|---|---|---|
| 0 | ST | 0 | -3 | 5 | CNA200 | -5.6 | -6.6 |
| 1 | PN | -3.7 | -6.0 | 6 | CNA500 | -6.6 | -6.6 |
| 2 | WNA | -4.8 | -6.0 | 7 | CNL200 | -3.7 | -4.7 |
| 3 | WNL | -3.7 | -4.9 | 8 | CNL500 | -3.7 | -3.7 |
| 4 | CN | -3.7 | -6.6 | | | | |

## DATA ANALYSIS

### Hass Effect Result

Using the data collected from 40 volunteers, the average time of identifying 2 different sounds $T_H$ equals 9.80 s. Since people take time to respond to the sound, the $T_H$ should represent the 9 s' group, taking the int part. As arranged, the group shows at 9th second is the 10th group, whose internal time is 32.5 ms, which matchs the classic result that people can identify two identica sound (same $L_s$ and frequency component) when their interval time is longer than 30 ms. Since the interval time is a relative parameter, this result has verified the online psychoacoustics test that it measures the relative data in some extent.

### Working Performance under Different Noises

In order to figure out their performance, the 16 average scores, which represent the average working performance (time and accuracy) under 8 different noises, were calculated from 40 volunteers' data. To normalize these average times, the total average time was calculated and was divided from 8 separate average times, based on which the time ratios under 8 different noises were calculated. The accuracy ratios were also calculated from 8 separate accuracies under 8 different noises in the same way. The normalization individual variations were summarised.

As shown in figure 4(a), it is obvious that the accuracy ratio has the opposite variation trends with the time ratio except the change with No.4 and No.5. To enhance the performance, time ratios were subtracted from the accuracy ratios, which were defined as performance ratio, was plotted in figure 4(b).

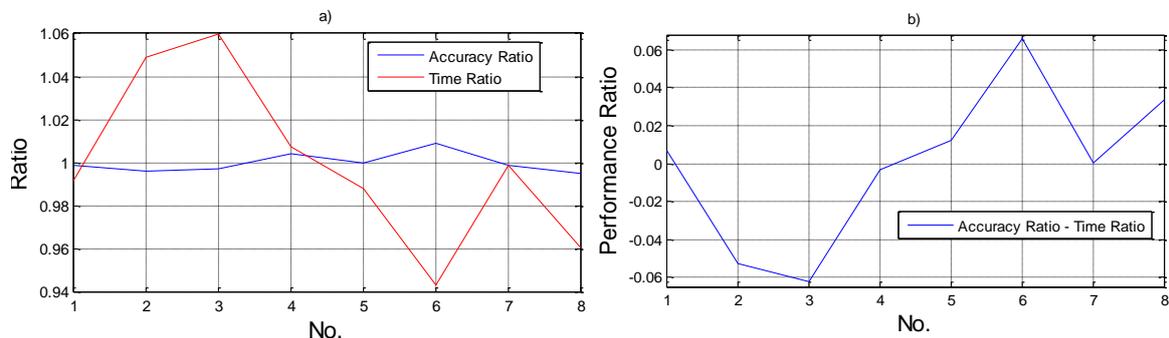

**Figure 4:** the working performance, including Accuracy Ratio, Time Ratio and Performance Ratio, under different kind of noise

No. 1 to 3 will be analyzed firstly. It is clear that the working performance under the pink noise enviverments is higher than the one under the white noise with identical $L_{s,A}$ and $L_{s,L}$, which means that the $L_{s,A}$ or $L_{s,L}$ could not represent their influence, the same evidence could be found among No.4, 5 and 6, and 4, 7 and 8. Since they have the same $L_{s,A}$ or $L_{s,L}$, the white noise with a much higher $L_s$ when frequency is higher than 3150 Hz, which is the most sensitive frequency to human being, also does harm to human performance. This result shows that either $L_{s,L}$ or $L_{s,A}$ could not present the influence of the noise to working performance individually, which is the flaw in the GB. Focus on No. 2 and 3, since they have the same frequency component and noise No. 2 is 1.1dB lower than noise No. 3, the working performance of No. 2 is slightly higher than No. 3, which has been shown in figure 4. It verifies the accuracy of our result because higher noise means worse working performance between 60 dB to 70 dB, and the same evidence can also be found between performance ratio No. 5 and 7, and No. 6 and 8, whose $L_s$ differences are 1.9 dB and 2.9 dB, respectively.

Secondly, focus on No. 4 to 8. Since No. 5 and 6 noises are consisted of No. 4 noise with 200 Hz and 500 Hz high pass filters respectively, the low frequency component of No. 4 is more abundant than that of No.5 and 6. Because of the same $L_{s,A}$, the $L_s$ of higher pitch of noise No.5 and 6 is higher than that of No.4. The high pass filter, distributes over the frequency higher than 200 Hz (0.1 dB higher in the left 19 bands) and 500 Hz (0.5 dB higher in the left 15 bands). So the $L_s$ differences on higher pitch among noise No.4 to 6 are quite small that people could not notify. And their difference mainly appears on the low frequency. Demonstrated from figure 4(b), the working performance of No. 5 and 6 are all higher than No.4. The same phenomenon could be found among No 7, 8 and 4.

The above analyses are based on the average data. Since the equipment among each volunteers was different, especially the frequency response of each earphone and sound card, the individual data may also be valuable. Every 2 scores among the 8 of individual data were compared and the comparisons results are shown in figure 5. In figure 5, (6>=7) means the Performance Ratio under noise No. 6 is no lower than the one under noise No. 7.

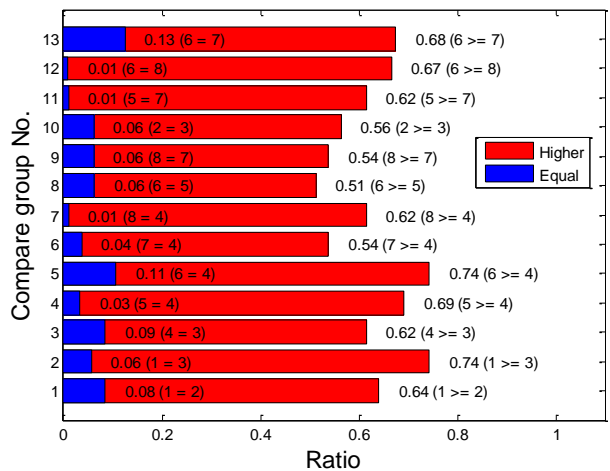

**Figure 5:** Result of Individual data comparison. The red bar present the percentage of No A higher that the one of No B (A and B shown in braket). The blue bar present the two percentages are equal.

Even the situations vary among each individual volunteer, at least more than 50%, 9 groups more than 62% of the relationships match that relationships conclude from the analyses of average data. What's more, focusing on group 4 and 5 in the comparisons, nearly more than 70% individual data shows that people have worse working performance under noise No.4 than that under noise No.5 and No.6. And the only difference among Noise No. 4, 5 and 6 is the frequency component that noise No.4 has more low frequency energy than any other noise. All these evidences indicate that $N_{LF}$ (25 Hz to 200 Hz) definitely does special harm to human working performance.

## CONCLUSION

In this paper, the low frequency noise standards and recommendations around the world are summarized. The European countries as well as Japan have one step ahead in the low frequency noise limitation, which is quite meticulous from the infrasound to 200 Hz. Compared to them, most of the Chinese National Standards still use A-weight sound pressure level as the limitation unit, which has been proved to have flaws in many studies, including the online psychoacoustic test conducted in this paper. Although Chinese National Standards begin to use low frequency noise limitation since 2008, it uses octave band and the limitation is higher than that in any other countries. Taiwan used A-weight sound pressure level to limit their low frequency noise from 20 Hz - 200 Hz as well. On the other hand, the online test has been proved useful in this paper via the Hass effect test as well as the low frequency noise test itself. Moreover, by requiring volunteers to participate in the colour recognition task under different controlled noise, it can be proved that low frequency noise does impact human working performance and A-weight and linear sound pressure levels could not present it perfectly.

## AKNOWLEAGMENT